\documentclass{blois}

\def\be{\begin{equation}}
\def\ee{\end{equation}}
\def\bea{\begin{eqnarray}}
\def\eea{\end{eqnarray}}

\newcommand{\Photo}{}

\begin{document}
\vspace*{3.5cm}
\title{FIRST CRYOGENIC TESTS ON BINGO INNOVATIONS}

\author{A.~ARMATOL$^{1}$, C.~AUGIER$^{2}$, D.~BAUDIN$^{1}$, G.~BENATO$^{3}$, J.~BILLARD$^{2}$, P.~CARNITI$^{4}$, M.~CHAPELLIER$^{5}$, A.~CHARRIER$^{6}$, F.~DANEVICH$^{7}$, M.~DE~COMBARIEU$^{6}$, M.~DE~JESUS$^{2}$, L.~DUMOULIN$^{5}$, F.~FERRI$^{1}$, J.~GASCON$^{2}$, A.~GIULIANI$^{5}$, H.~GOMEZ$^{1}$, C.~GOTTI$^{4}$, PH.~GRAS$^{1}$, M.~GROS$^{1}$, A.~JUILLARD$^{2}$, H.~KHALIFE$^{1}$, V.V.~KOBYCHEV$^{7}$, M.~LEFEVRE$^{1}$, P.~LOAIZA$^{5}$, S.~MARNIEROS$^{5}$, PH.~MAS$^{1}$, E.~MAZZUCATO$^{1}$, J.F.~MILLOT$^{1}$, C.~NONES$^{1}$, G.~PESSINA$^{4}$, D.V.~PODA$^{5}$, J.A.~SCARPACI$^{5}$, O.~TELLIER$^{1}$, V.I.~TRETYAK$^{7}$, M.M.~ZARYTSKYY$^{7}$, A.~ZOLOTAROVA$^{1}$}

\address{$^{1}$IRFU, CEA, Universit\'e Paris-Saclay, Saclay, France \\
$^{2}$Institut de Physique des 2 Infinis, Lyon, France \\
$^{3}$INFN Laboratori Nazionali del Gran Sasso, Assergi (AQ), Italy \\
$^{4}$INFN Sezione di Milano - Bicocca, Milano, Italy \\
$^{5}$Universit\'e Paris-Saclay, CNRS/IN2P3, IJCLab, Orsay, France\\
$^{6}$IRAMIS, CEA, Universit\'e Paris-Saclay, Saclay, France \\
$^{7}$Institute for Nuclear Research of NASU, Kyiv, Ukraine 
}

\maketitle\abstracts{
Neutrinoless double-beta decay ($0\nu2\beta$) is a hypothetical rare nuclear transition. Its observation would provide an important insight about the nature of neutrinos (Dirac or Majorana particle) demonstrating that the lepton number is not conserved. BINGO (Bi-Isotope $0\nu2\beta$ Next Generation Observatory) aims to set the technological grounds for future bolometric $0\nu2\beta$ experiments. It is based on a dual heat-light readout, i.e. a main scintillating absorber embedding the double-beta decay isotope accompanied by a cryogenic light detector. BINGO will study two of the most promising isotopes: $^{100}$Mo embedded in Li$_2$MoO$_4$ (LMO) crystals and $^{130}$Te embedded in TeO$_2$. BINGO technology will reduce dramatically the background in the region of interest, thus boosting the discovery sensitivity of $0\nu2\beta$. The proposed solutions will have a high impact on next-generation bolometric tonne-scale experiments, like CUPID. In this contribution, we present the results obtained during the first tests performed in the framework of BINGO R\&D.}

\section{Introduction} \label{sec:one}

The current largest bolometric experiment CUORE, installed in LNGS (Italy), is taking data since 2017 using TeO$_2$ crystals as bolometers\cite{CUOREres}. However, CUORE is expecting around 50 background events per year, mainly due to $\alpha$ particles coming from surface radioactivity\cite{CUOREbckg}. To quantify the background of an experiment, we use a parameter called the background index, \textit{b}, that corresponds to the number of expected background events per kilogram of detector per keV per year (ckky) in the region of interest (ROI). For CUORE, $b$ is of the order of 10$^{-2}$ ckky limitating the sensitivity to the Majorana effective mass\cite{DBDreview} (m$_{\beta\beta}$) in case of light Majorana neutrino exchange mechanism. For the next-generation experiment CUPID\cite{CupidCDR}, the goal is to increase the sensitivity to fully explore the inverted hierarchy region\cite{DBDreview}. CUPID will use scintillating bolometers and the dual heat-light readout allowing to reject $\alpha$ events and moving to $^{100}$Mo as $\beta\beta$ isotope ($Q_{\beta\beta}$ above the natural $\gamma$ radioactivity endpoint). The CUPID-Mo\cite{CupidMo} experiment has proved the good performance of LMO crystals and has obtained a rejection superior at 99\% for $\alpha$ background, allowing CUPID to reach $b\sim$10$^{-4}$ ckky using the same crystals. However, neutrino oscillation experiments tend to point to the normal neutrino mass hierarchy\cite{OscN}. BINGO takes place in this framework: preparing the exploration of the normal hierarchy region for the next-to-next generation experiment by reducing drastically the number of background events in the ROI down to $b\sim$10$^{-5}$ ckky.
In addition to the use of $^{100}$Mo and $^{130}$Te, BINGO proposes three main improvements to the current state of the art: a revolutionary detector assembly, a cryogenic active veto and light detectors exploiting Neganov-Trofimov-Luke (NTL) effect\cite{NL_LD,NL2}. We will focus here on the two first ones.
The revolutionary detector assembly will minimize the amount of passive material surrounding the detectors. We propose to only have a small copper structure on which two crystals are held by nylon wires with light detectors placed between the copper and the absorbers to shield them from it. This assembly will reduce the contribution of the surface radioactivity to the background. We have done a test on a first prototype using small LMO crystals that will be described later in this article. We present also the use for the first time in bolometric experiments of a cryogenic active veto. It will be composed of scintillators, ZnWO$_4$ (ZWO) or Bi$_4$Ge$_3$O$_{12}$ (BGO) crystals surrounding the detector area. By reading only the scintillation light, thanks to two NTL light detectors, one should be able to reject most of the external $\gamma$ background coming from the environment and reduce the internal $\gamma$ and $\beta$ background due to residual contamination of the detector-array structure. It is essential since for example the expected $0\nu2\beta$ peak of $^{130}$Te ($Q_{\beta\beta}$=2527.5 keV) is in a region with a rather high amount of $\gamma$ events. We have performed first tests with the two candidates in order to have a preliminary comparison between them and the results are presented below.

\section{Test of the first BINGO prototype with the nylon assembly}\label{subsec:2}

\begin{figure}[h]
\begin{minipage}{0.24\linewidth}
\centerline{\includegraphics[width=1\linewidth]{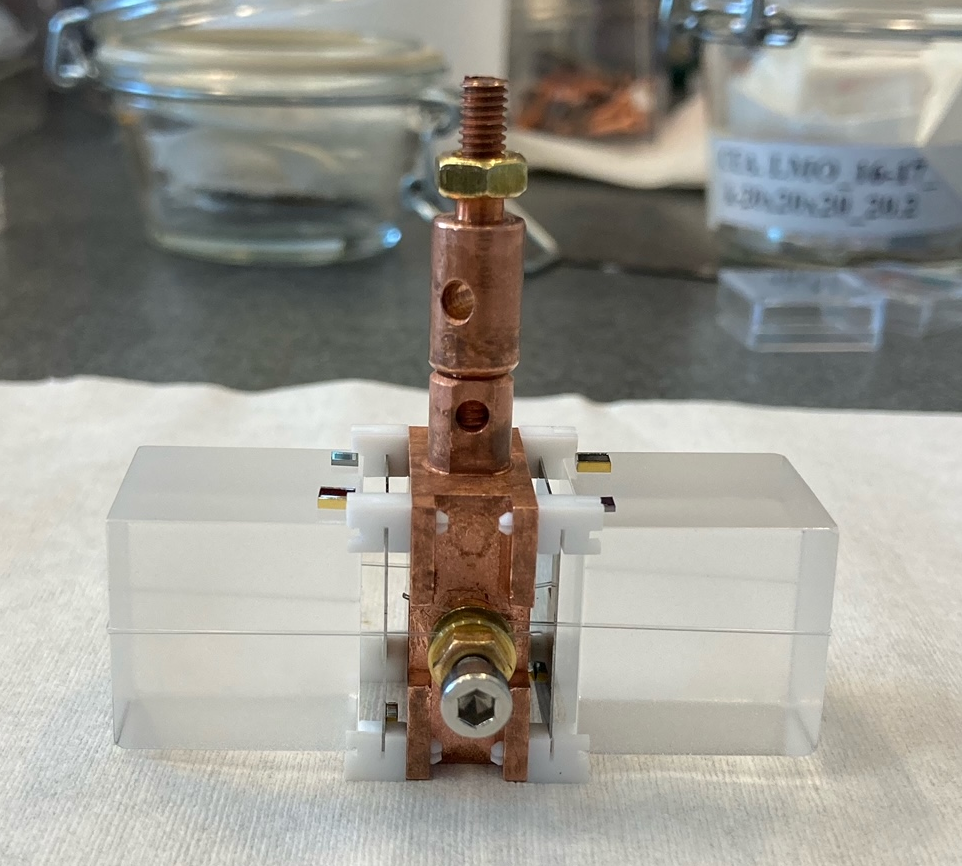}}
\end{minipage}
\hfill
\begin{minipage}{0.42\linewidth}
\centerline{\includegraphics[width=1\linewidth]{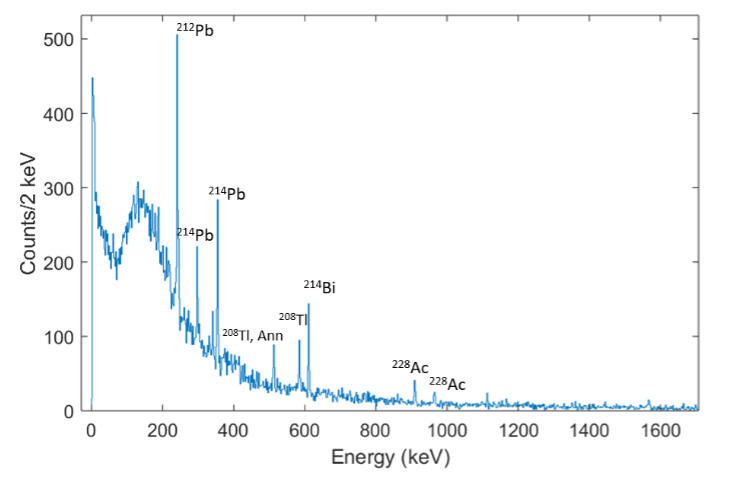}}
\end{minipage}
\hfill
\begin{minipage}{0.32\linewidth}
\centerline{\includegraphics[width=1\linewidth]{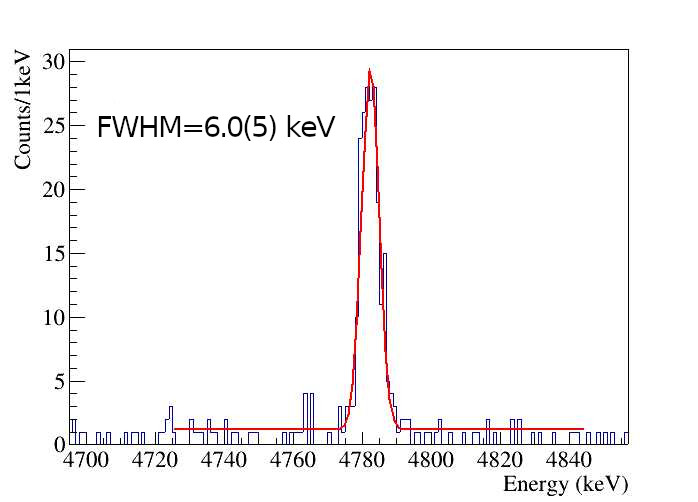}}
\end{minipage}
\caption[]{a) First prototype of the nylon assembly (left). b) Energy spectrum of a Th calibration for LMO4 (center). c) $^{6}$Li(n,t)$\alpha$ peak obtained during the same measurement (right).}
\label{fig:nylon}
\end{figure}

We have assembled a first prototype of the nylon assembly to demonstrate the feasibility of the technology proving that this way of holding the crystals and the light detectors is efficient even at low temperature. The main concerns were the possible introduction of mechanical vibrations, on which the performance of bolometers depend a lot. We have used 2 small natural LMO crystals ($20\times20\times20$ mm$^3$), called LMO2 and LMO4, each equipped with an NTD Ge sensor glued on one of the faces using 9 spots of Araldite epoxy glue. A heater for temperature stabilization was also glued on the same face. For light detectors, we used 2 Ge square wafers ($20\times20\times0.25$ mm$^3$) with a Ge NTD glued on it with 3 glue spots. To avoid the crystal to touch the copper holder, 4 PTFE pieces were placed at each corner between them. They had in the middle a slot to host the light detector in between and then "sandwich" it when a sufficiently high force was applied by the nylon wire on the crystal. We used a weight of 2.5 kg to fix the nylon wire tension for each crystal. We made sure that they were passing in the middle of the crystal faces before blocking them. At the end we have obtained an assembly really rigid with the detectors well-kept in place (Fig.~\ref{fig:nylon}a). The prototype was then put inside a closed structure made of copper with the inner part covered with reflective foil before being placed in a dilution cryostat equipped with cryogenic suspensions.
 
We have done several measurements at different temperatures, the results presented here are the best obtained at 15 mK where a Th wire was attached to the cryostat in order to make easier the calibration. 
The data were processed using an analysis tool applying the Gatti-Manfredi Optimum Filter\cite{GatMan} to the stream and therefore optimising the reconstructed energy resolution. In general, the two crystals have shown excellent performance. We have obtained high sensitivities for both crystals, which are good for bolometers of this size and the baseline FWHM were below 1 keV (Table~\ref{tab:resultsnylon}). 
\begin{table}[]
\caption[]{Best performance at 15 mK of the detectors during the prototype test of the nylon assembly}
\label{tab:resultsnylon}
\vspace{0.4cm}
\begin{center}
\begin{tabular}{|c|c|c|c|c|c|}
\hline
Detector & $R_{NTD}$  & Sensitivity   & FWHM$_{noise}$ & FWHM at 609 keV & FWHM at $^{6}$Li(n,t)$\alpha$ \\
    & (M$\Omega$) & (nV/keV) & (keV) & (keV) & (keV) \\
\hline
LMO2 & 3.2 & 178 & 0.95 & 2.3 & 8.5 \\
\hline
LMO4 & 4.5 & 300 & 0.85 & 2.4 & 6.0 \\
\hline
\end{tabular}
\end{center}
\end{table}
The energy resolutions obtained on Th peaks are promising as well (Fig.~\ref{fig:nylon}b). Moreover, for LMO4, we have measured the best FWHM for the peak of the $\alpha$ coming from the $^{6}$Li neutron capture ever measured in an above-ground measurement (Fig.~\ref{fig:nylon}c).
In general, this first prototype test has allowed us to prove that nylon wires are a suitable way to hold both crystals and light detectors to a small copper structure, without preventing them to obtain good performance as bolometers. This is really encouraging since it will allow us to reduce the amount of passive materials for future bolometric experiments. A new test is on going using bigger LMO crystals (45$\times$45$\times$45 mm$^3$) in an improved assembly to confirm the high performance of the detectors set-up.

\section{First comparison between the two veto candidates}

\begin{figure}[h]
\begin{minipage}{0.29\linewidth}
\centerline{\includegraphics[width=0.65\linewidth]{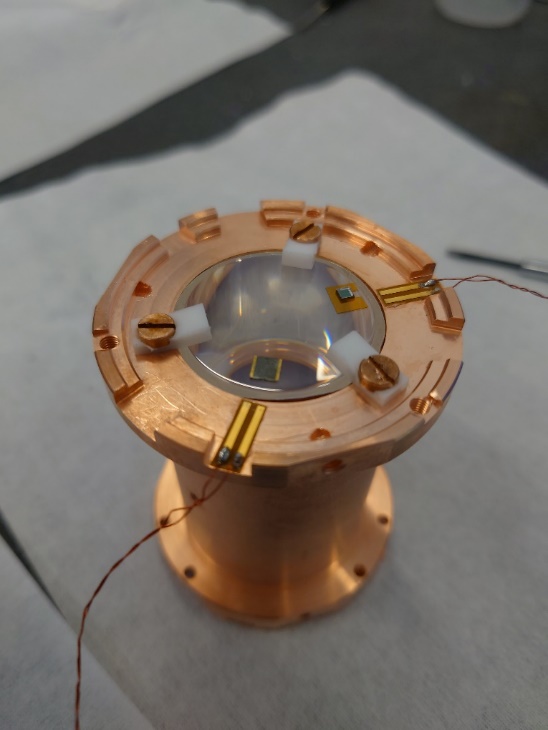}}
\end{minipage}
\hfill
\begin{minipage}{0.27\linewidth}
\centerline{\includegraphics[width=0.8\linewidth]{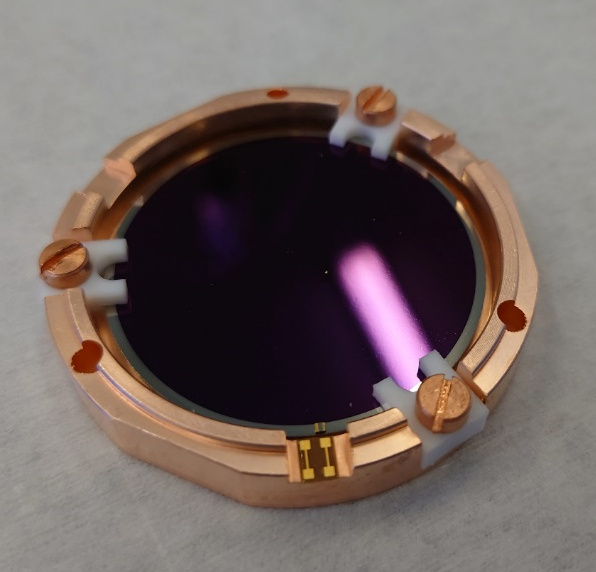}}
\end{minipage}
\hfill
\begin{minipage}{0.42\linewidth}
\centerline{\includegraphics[width=1\linewidth]{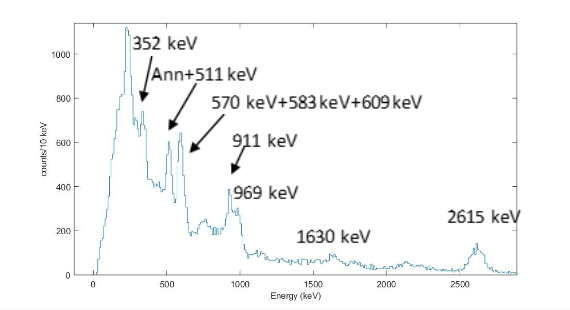}}
\end{minipage}
\caption[]{a) BGO crystal placed inside the copper holder (left). b) The SiO coated Ge light detector mounted inside the cover (center). c) Energy spectrum of $\gamma$-ray quanta from environmental radioactivity and contamination of the BGO crystal by $^{207}$Bi measured by the BGO detector using scintillation readout (right).}
\label{fig:vetoassemb}
\end{figure}

In order to select the best compound between BGO and ZWO as crystal for the BINGO cryogenic veto we have performed a first test at low temperature. We made a run for each candidate using the same crystal size and shape (cylindrical $\oslash$30$\times$60 mm$^2$) and placing them in a copper holder.
Although the veto crystals will not be used as bolometers in the final BINGO configuration, we decided for the two tests to glue one NTD and one heater on the crystal to have the possibility to look at the heat channel in addition to the light channel. The crystal (BGO or ZWO) was resting on 3 PTFE pieces on the bottom of the holder and blocked by 3 other pieces on the top (Fig.~\ref{fig:vetoassemb}a). All the inner part of the copper structure was covered with reflective foil to enhance light collection. The scintillation light was measured with the same SiO coated Ge light detector with an NTD glued on it for both crystals that was mounted inside the cover of the full assembly and held by PTFE clamps (Fig.~\ref{fig:vetoassemb}b). It was placed at the top of the full structure. The tests were performed in the same cryostat than the one used for the nylon assembly.

We have done long background measurements at 15 mK for both crystals with the same procedure to analyze the data as the one already mentioned in Section \ref{subsec:2}. In both tests, the light detector showed excellent performance with a high sensitivity and a low baseline noise, allowing us to really compare the results on the light yield for the two crystals (Table~\ref{tab:resultsveto}). Concerning the BGO, it is known that this crystal is extremely slow to cool down\cite{BGO} and indeed we have confirmed this. The BGO NTD has never reached a resistance high enough to be readout even after two weeks of data taking. Therefore, we only had at our disposal the light measurement from the Ge auxiliary bolometer. However, the light signal was enough to well reconstruct the spectrum of the $\gamma$-ray quanta absorbed in the crystal  (Fig.~\ref{fig:vetoassemb}c). We performed a calibration of the light detector using the muons bump and computed the light yield with the $\gamma$ peaks visible in the spectrum. We have obtained a really high light yield of 28.0 keV/MeV (Table~\ref{tab:resultsveto}).
\begin{table}[]
\caption[]{Performance at 15 mK of the Ge light detector in the two configurations}
\label{tab:resultsveto}
\vspace{0.4cm}
\begin{center}
\begin{tabular}{|c|c|c|c|c|c|}
\hline
Crystal & $R_{NTD}$  & Sensitivity & FWHM$_{noise}$ & Light yield & Energy threshold \\
    & (M$\Omega$) & ($\mu$V/keV) & (eV) & (keV/MeV) & (keV) \\
\hline
BGO & 2.2 & 1.77 & 127 & 28.0 & 10 \\
\hline
ZWO & 1.7 & 1.73 & 146 & 13.6 & 25 \\
\hline
\end{tabular}
\end{center}
\end{table}
On the other hand, the ZWO crystal cooled down normally to 15 mK, the heat channel was then available. We measured a light yield of 13.6 keV/MeV, almost the half of the BGO one. However, this was enough to reconstruct the spectrum of the $\gamma$-ray quanta absorbed in the crystal using only the scintillation light. We did also a first estimation of the energy threshold for each detector using a value corresponding to 5$\sigma$ of the baseline assuming a Gaussian distribution and without taking into account the detection efficiency. The obtained values need to be investigated with deeper studies. In conclusion, both candidates have shown satisfying performance to be used as the BINGO veto; it is too early to make a choice and more tests and simulations are required. For example, we know that a $^{207}$Bi contamination is usually present inside BGO\cite{BGOcont}, so we need to evaluate how harmful this background contribution could be. Moreover, we have to confirm that it is possible to grow ZWO crystals in the size needed for the final veto (a length of 30 cm).

\section*{Acknowledgments}
The BINGO project has received funding from the European Research Council (ERC) under the European Union’s Horizon 2020 research and innovation program (grant agreement No 865844). The INR NASU group was supported in part by the National Research Foundation of Ukraine Grant no. 2020.02/0011. We would like to thank also Ph.~Forget and the mechanical workshop of SPEC for their valuable help.

\end{document}